\documentclass[11pt,a4paper,twocolumn]{article}
\usepackage[a4paper,margin=1in,columnsep=0.25in]{geometry}
\usepackage{times}
\usepackage{latexsym}
\usepackage[T1]{fontenc}
\usepackage[utf8]{inputenc}
\usepackage{microtype}
\usepackage{inconsolata}
\usepackage{graphicx}
\usepackage{amsmath,amssymb}
\usepackage{booktabs}
\usepackage{multirow}
\usepackage{url}
\usepackage{xcolor}
\usepackage{array}
\usepackage{tabularx}
\usepackage{placeins}
\usepackage[round,authoryear]{natbib}
\usepackage[colorlinks=true,linkcolor=blue,citecolor=blue,urlcolor=blue]{hyperref}
\newcommand{\method}{BReF}

\newcommand{\prob}{\mathbf{p}}
\newcommand{\resp}{\mathbf{r}}
\newcommand{\figplaceholder}[2]{%
\fbox{\begin{minipage}[c][#1][c]{0.94\linewidth}\centering\small #2\end{minipage}}}

\newcommand{\probesensitivityfigurepath}{figures/bref_probe_sensitivity.png}
\newcommand{\benchmarkfigurepath}{figures/bref_benchmark_protocol.png}
\newcommand{\pipelinefigurepath}{figures/bref_pipeline.png}
\newcommand{\heatmapfigurepath}{figures/bref_exact_parent_heatmap.png}
\newcommand{\ksensitivityfigurepath}{figures/bref_k_sensitivity.png}
\newcommand{\perturbcoveragefigurepath}{figures/bref_perturbation_coverage.png}

\title{\textbf{Beyond QA Matching: Perturbation-Response Fingerprinting via Probability Distributions for Large Language Models}}
\author{Jichao Zeng$^1$, Yanli Chen$^1$ and Hanzhou Wu$^{1,2}$~\footnote{Corresponding author: \emph{Dr. Hanzhou Wu}}\\
$^1$Guizhou Normal University\quad $^2$Shanghai University}
\date{}

\begin{document}
\raggedbottom
\maketitle

\begin{abstract}
Large language models are often instruction-tuned, specialized, quantized, or otherwise transformed, making fine-grained provenance difficult. In this paper, we introduce \method{}, a training-free fingerprint that compares how probability distributions over four answer-option labels A/B/C/D move under controlled textual perturbations. For each pair of models, \method{} selects 25 jointly responsive probes and compares their perturbation log-ratio (PLR) response directions by global cosine similarity. On a unified benchmark with 34 checkpoints, 22 documented direct-parent relations, and 411 suspect--candidate pairs, \method{} retrieves the documented parent in 22/22 cases (MRR=1.0000), with DP--DF AUC 1.0000. Same-family discrimination is harder (DP--SF AUC 0.8969), and paired tests show a significant exact-retrieval gain over a magnitude-only Top-25 control. Together with static, random-probe, permutation, calibration, and transformation-level controls, the results show that strong pooled separation does not guarantee correct parent ranking among closely related checkpoints, verifying the superiority of our work.
\end{abstract}

\section{Introduction}

Large language models (LLMs) are rarely deployed as immutable pretrained checkpoints. A base model may be instruction-tuned, specialized for code or mathematics, quantized for deployment, or otherwise post-trained before users interact with it. Distinguishing such closely related checkpoints is difficult when only model outputs and their probability distributions are available. This motivates behavioral techniques for fine-grained model identification and provenance screening.

A straightforward behavioral fingerprint compares final responses to shared prompts; for multiple-choice inputs, this reduces to hard answer agreement. Throughout this paper, A, B, C, and D denote only the four answer-option labels. Hard agreement discards probability information: two models may choose the same option with different uncertainty, while post-training may change the argmax without eliminating characteristic perturbation-response behavior. Prior fingerprinting methods therefore exploit richer behavioral or representation-level signals \citep{cao2021ipguard,guan2022sac,peng2022uap,bai2024ibsf,pasquini2025llmmap,zhang2025reef,shao2026zeroprint}.

We investigate a probability-access setting in which the four designated option probabilities, or sufficient option log-probabilities to normalize them into a four-way probability vector, are available for each multiple-choice query. Section~\ref{sec:problem_setting} formalizes this observation protocol.

\begin{figure}[t]
\centering
\IfFileExists{\probesensitivityfigurepath}{%
    \includegraphics[width=0.98\linewidth]{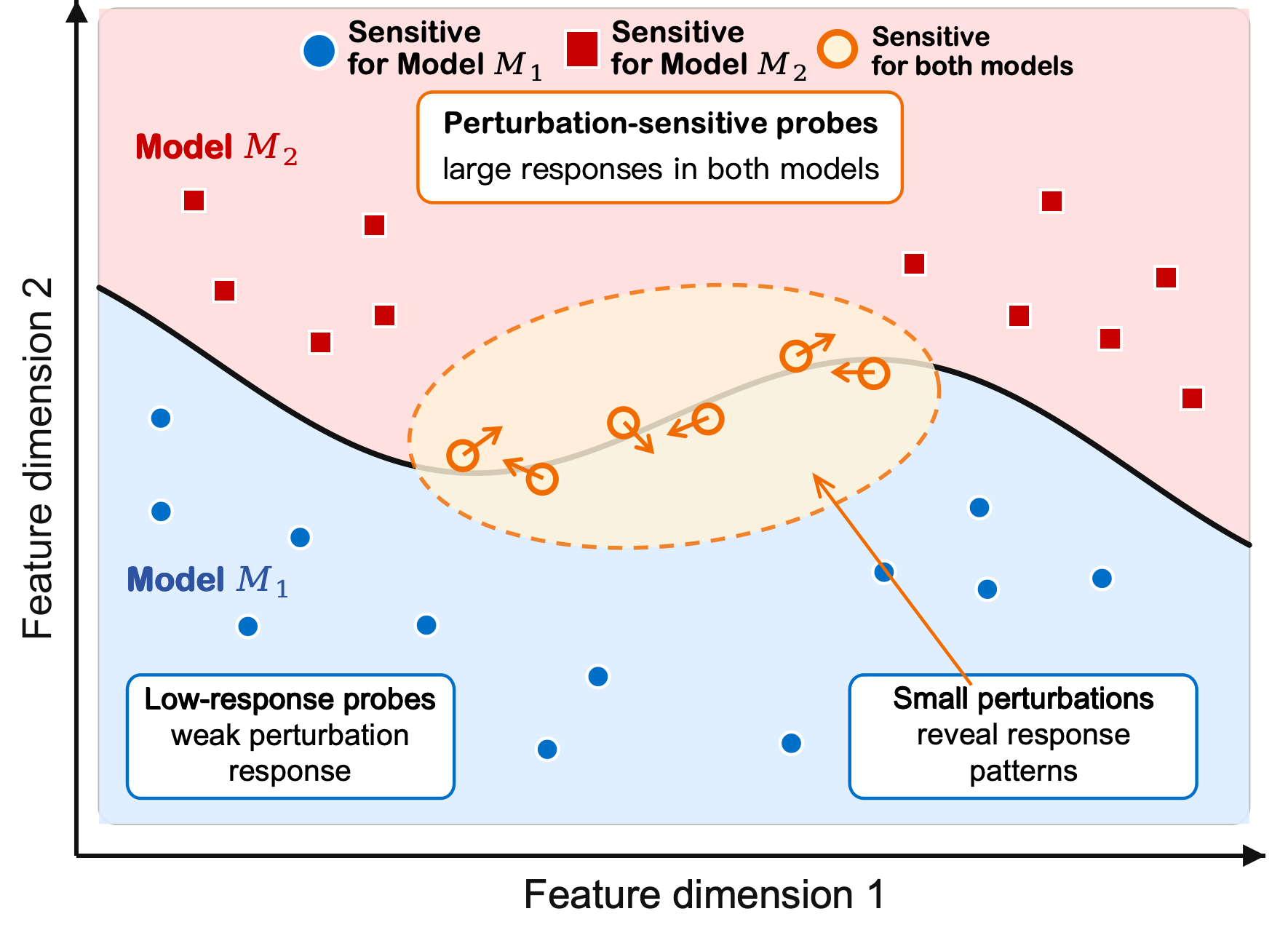}%
}{%
    \figplaceholder{1.65in}{%
    \textbf{PROBE-SENSITIVITY FIGURE FILE NOT FOUND.}\\
    Expected path: \texttt{figures/bref\_probe\_sensitivity.png}.}%
}
\caption{Pair-specific perturbation-response sensitivity. Blue and red points denote probes with large perturbation responses for the compared models $M_1$ and $M_2$, respectively. Their overlap represents jointly responsive probes favored by the geometric joint magnitude described in Eq.~(\ref{eq:joint}).}
\label{fig:probe_sensitivity}
\end{figure}

Our key design principle is simple: \textbf{response magnitude chooses where to compare, while response direction determines similarity}. BReF represents each probe by a 52-dimensional perturbation log-ratio (PLR) response vector, selects $K=25$ probes that are jointly responsive for the compared model pair, and compares their concatenated response directions using global cosine similarity. Probe selection uses response magnitude only, whereas the final score uses response direction; neither stage uses family or provenance labels.
Figure~\ref{fig:probe_sensitivity} illustrates the jointly responsive probe region favored by the geometric joint magnitude in Eq.~(\ref{eq:joint}).

Our contributions are:
\begin{itemize}
    \item We introduce a probability-distribution fingerprint based on perturbation log-ratio responses, pair-specific joint response selection, and global cosine similarity.
    \item We formulate a role-constrained direct-parent retrieval benchmark with 34 checkpoints, 22 documented direct-parent relations, a shared candidate registry, and 411 suspect--candidate pairs.
    \item We compare against REEF, MET, ZeroPrint, LLMmap, and a matched QA-agreement baseline on the same direct-parent retrieval benchmark and candidate registry, and show statistically significant gains after multiple-comparison correction.
    \item We systematically validate the mechanism through static-probability,
    energy-only, selector, probe-count, perturbation-count,
    semantic-option-permutation, and probability-calibration controls,
    together with a tokenizer audit.
\end{itemize}

\section{Related Work}
\paragraph{Boundary and output-structure fingerprints.}
IPGuard fingerprints DNN classification boundaries without modifying the victim model \citep{cao2021ipguard}. SAC instead exploits correlations among model outputs \citep{guan2022sac}, while universal-adversarial-perturbation fingerprinting characterizes global decision-boundary structure \citep{peng2022uap}. IBSF constructs samples near intersections of decision boundaries using entropy-sensitive objectives and performs tamper detection from top-1 labels \citep{bai2024ibsf}. Together, these studies motivate the use of richer behavioral structure beyond ordinary hard-label agreement. BReF differs in operating on perturbation-induced LLM option-probability responses and in selecting perturbation-sensitive probes separately for each compared pair.

\paragraph{LLM fingerprinting and API auditing.}
Existing LLM fingerprinting methods span substantially different access assumptions. Instructional Fingerprinting embeds a secret instruction key through lightweight tuning \citep{xu2024instructional}, while HuRef derives owner-side fingerprints from model parameters \citep{zeng2024huref} and UTF exploits under-trained tokens as identification cues \citep{cai2025utf}. Non-invasive verification has also been studied through pooled membership inference \citep{wu2022pmi}. Under inference-time access, LLMmap identifies model versions from generated responses \citep{pasquini2025llmmap}, REEF compares internal representations using centered kernel alignment \citep{zhang2025reef}, and ZeroPrint estimates response gradients from black-box generations \citep{shao2026zeroprint}. Recent work has further investigated sensitive fingerprint samples \citep{bai2025esf}, merge-resistant fingerprints \citep{yamabe2025mergeprint}, concept-response fingerprints \citep{tang2026cref}, and robustness-oriented fingerprinting \citep{zhao2026robust}. Model Equality Testing (MET) instead formulates API auditing as two-sample testing between output distributions \citep{gao2025met}, while a recent SoK surveys the broader LLM-fingerprinting landscape \citep{shao2025sok}. Fingerprinting has also been investigated for API-protected LLMs \citep{yang2026hidden}. BReF occupies a different operating point: it requires no model modification or hidden-state access, but assumes access to the four designated option probabilities and compares their perturbation-induced response geometry for pairwise model ranking.

\paragraph{Multiple-choice probability protocols.}
Reading next-token probabilities for A/B/C/D is sensitive to tokenization and answer formatting, and seemingly minor choices can alter MCQA accuracy and model rankings \citep{sanzguerrero2025mind}. BReF therefore fixes a common four-label probability protocol and evaluates semantic option permutations as an end-to-end robustness control.

\section{Problem Formulation}
\label{sec:problem_setting}
Let $M$ be an LLM and $x_i$ a four-option multiple-choice probe. We use $\mathrm{A}$, $\mathrm{B}$, $\mathrm{C}$, and $\mathrm{D}$ exclusively to denote the four answer-option labels; model identifiers are denoted separately by $M$, $M_1$, and $M_2$. Let $\mathcal{Y}=\{\mathrm{A},\mathrm{B},\mathrm{C},\mathrm{D}\}$ denote the answer-label set. For condition $t$, we observe the option-probability vector
\begin{equation}
\prob_{M,i,t}=[p_{\mathrm{A}},p_{\mathrm{B}},p_{\mathrm{C}},p_{\mathrm{D}}]\in\Delta^3,
\end{equation}
where $\Delta^3$ is the four-way probability simplex. If option log-probabilities $\ell_c$ are available instead, we normalize only the four designated answer labels,
\begin{equation}
p_c=
\frac{\exp(\ell_c)}
{\sum_{d\in\mathcal{Y}}\exp(\ell_d)},
\; \forall c\in\mathcal{Y}.
\end{equation}
The method requires the complete four-option probability vector; interfaces that provide fewer than the four designated option values are outside the current protocol.

\paragraph{Probability observation protocol.}
The same prompt construction and answer-label convention are applied across all evaluated checkpoints. Robustness to tokenization, answer position, perturbation coverage, and probability calibration is evaluated later in the paper.

\begin{table}[t]
\centering
\footnotesize
\caption{Signals and objectives of different methods. ``Option prob.'' denotes direct A/B/C/D probabilities or sufficient option log-probabilities.}
\setlength{\tabcolsep}{2.5pt}
\renewcommand{\arraystretch}{1.14}
\begin{tabularx}{\columnwidth}{@{}l l c >{\raggedright\arraybackslash}X@{}}
\toprule
Method & Required signal & Modify? & Native objective \\
\midrule
LLMmap & text & No & version ID \\
MET & samples & No & model equality \\
ZeroPrint & text & No & fingerprint similarity \\
REEF & hidden repr. & No & derivation relation \\
Instr. FP & text & Yes & ownership key \\
\textbf{BReF} & \textbf{option prob.} & \textbf{No} & \textbf{pair similarity} \\
\bottomrule
\end{tabularx}
\label{tab:access}
\end{table}

\paragraph{Claim boundary.}
Given two models $M_1$ and $M_2$, BReF outputs a behavioral similarity score $S(M_1,M_2)\in[0,1]$. No family or lineage label enters the score. In direct-parent retrieval, a suspect model $Q$ is supplied together with a registry $\mathcal{P}=\{P_1,\ldots,P_m\}$ of candidate parent checkpoints. Each candidate $P_j$ is ranked by $S(Q,P_j)$, and retrieval is considered correct when the documented direct parent is ranked highest. The benchmark defines the suspect and candidate roles, while BReF provides behavioral evidence for ranking; the resulting score should not be interpreted as standalone proof of causal lineage or legal ownership.

\begin{figure}[t]
\centering
\IfFileExists{\benchmarkfigurepath}{%
\includegraphics[width=\columnwidth]{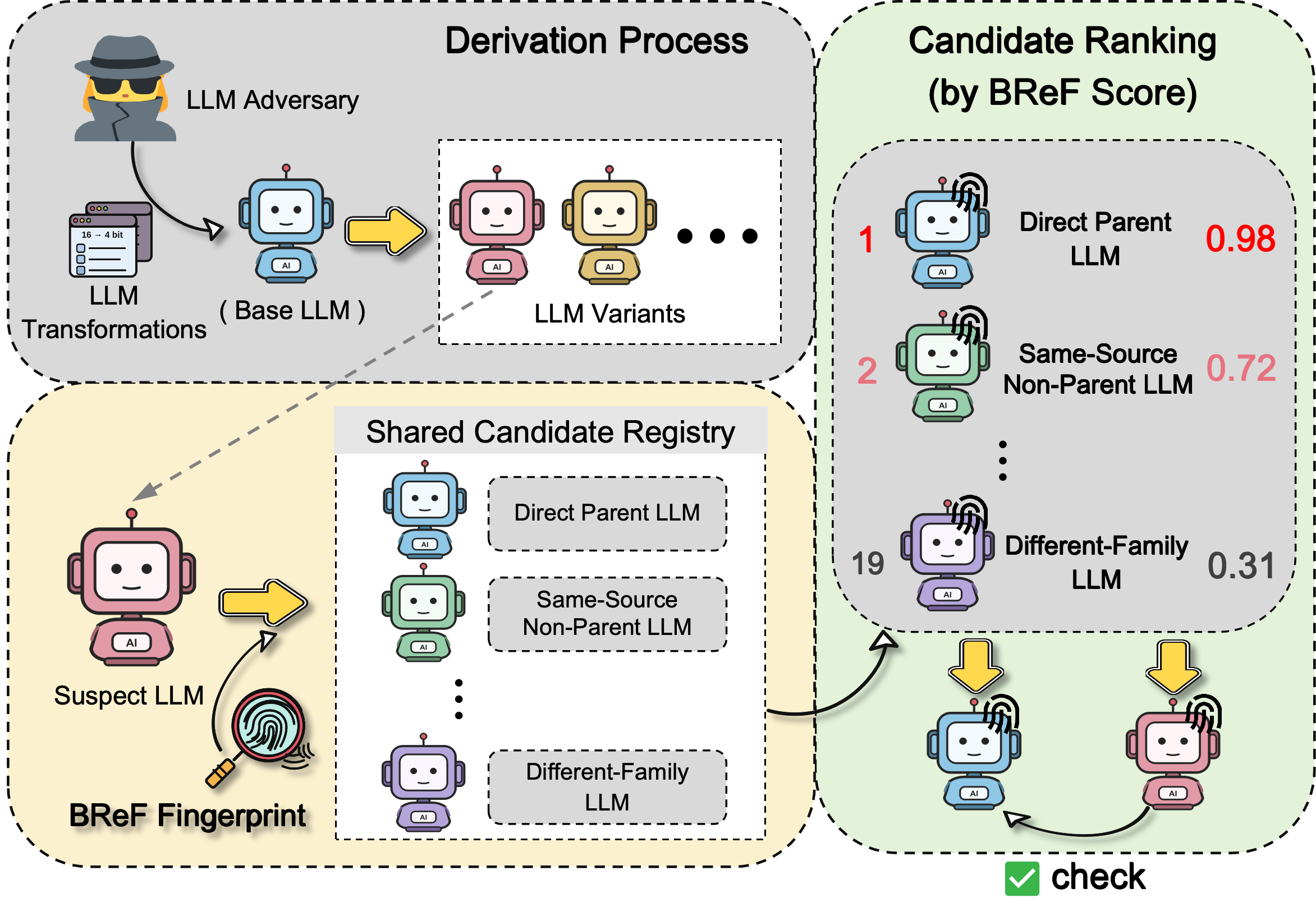}%
}{%
\figplaceholder{2.15in}{\textbf{BENCHMARK SCHEMATIC FILE NOT FOUND.}\\
Expected path: \texttt{figures/bref\_benchmark\_protocol.png}.}%
}
\caption{Unified direct-parent retrieval protocol. Benchmark provenance defines candidate roles; BReF scores and ranks the shared registry. Numerical scores are illustrative.}
\label{fig:benchmark}
\end{figure}

Figure~\ref{fig:benchmark} summarizes this retrieval protocol: benchmark provenance defines the suspect and candidate roles, while BReF supplies the scores used to rank the shared registry.

\section{BReF Method}
\begin{figure*}[t]
    \centering
    \IfFileExists{\pipelinefigurepath}{%
        \includegraphics[width=0.98\textwidth]{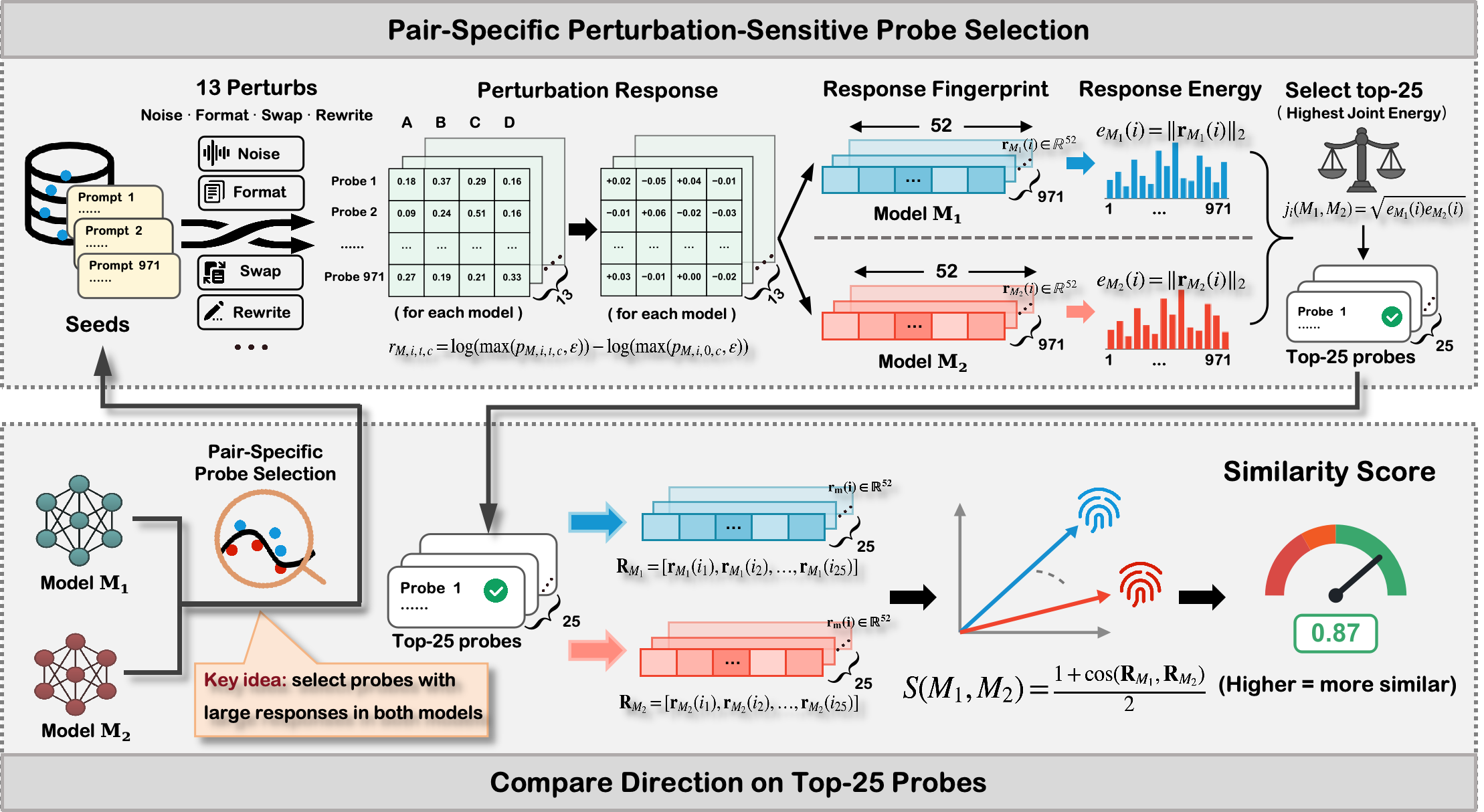}%
    }{%
        \figplaceholder{1.35in}{\textbf{PIPELINE FIGURE FILE NOT FOUND.}\\
        Expected path: \texttt{figures/bref\_pipeline.pdf}.\\[2pt]
        Upload the flowchart PDF to that path in Overleaf; the figure will be inserted automatically.}%
    }
    \caption{BReF pipeline. Thirteen fixed perturbations produce PLR responses; joint response magnitude selects Top-25 probes, and global cosine similarity compares their response directions.}
    \label{fig:pipeline}
\end{figure*}

Figure~\ref{fig:pipeline} summarizes the BReF scoring pipeline. The 13 fixed edits are shown schematically under four compact icon groups---Noise, Format, Swap, and Rewrite; the exact operations are listed in Appendix~\ref{app:probes}.

\subsection{Perturbation-response representation}
Let the fixed probe pool be $\mathcal{X}=\{x_1,\ldots,x_N\}$. The benchmark uses $N=971$ four-option probes, each evaluated under an unperturbed baseline $t=0$ and the fixed clean13 bank $t=1,\ldots,13$. The exact deterministic operations and probe-bank construction are documented in Appendix~\ref{app:probes}.

For model $M$, probe $i$, perturbation $t\ge 1$, and option $c$, define the clipped perturbation log-ratio (PLR) response
\begin{equation}
\begin{aligned}
r_{M,i,t,c}
&=\log\!\big(\max(p_{M,i,t,c},\varepsilon)\big)\\
&\quad-\log\!\big(\max(p_{M,i,0,c},\varepsilon)\big),
\end{aligned}
\label{eq:plr}
\end{equation}
with fixed $\varepsilon=10^{-12}$. Flattening all $T\times4$ entries yields $\resp_{M,i}\in\mathbb{R}^{52}$.

\paragraph{Interpretation of PLR.}
For a fixed probe and option, the PLR coordinate is $\log p^{(t)}-\log p^{(0)}=\log(p^{(t)}/p^{(0)})$ after clipping. Positive values indicate increased option probability under the perturbation and negative values indicate a decrease. This response-relative representation is evaluated against static-probability controls and calibration robustness experiments later in the paper.

\subsection{Pair-specific perturbation-sensitive probe selection}
For each probe, we compute the response magnitude
\begin{equation}
e_M(i)=\|\resp_{M,i}\|_2.
\end{equation}
For a model pair $(M_1,M_2)$, BReF defines the joint response magnitude
\begin{equation}
j_i(M_1,M_2)=\sqrt{e_{M_1}(i)e_{M_2}(i)}.
\label{eq:joint}
\end{equation}
The geometric mean is high only when \emph{both} models respond. Importantly, Eq.~(\ref{eq:joint}) uses only norms: it contains no dot product, cosine, label, or family information and therefore does not directly select probes because their response directions are already similar. We deterministically retain the $K=25$ probes with largest $j_i$, breaking ties by probe index. We denote this pair-specific set by $\mathcal{I}_{12}$.

The selection is pair-specific, so a fixed suspect may use different Top-25 subsets against different candidates. Every comparison nevertheless uses the same 971-probe pool, $K$, response representation, and similarity definition; only the identities of the selected probes may differ across pairs. Accordingly, BReF defines a pair-conditioned behavioral fingerprinting score rather than a single candidate-independent embedding assigned to
each model.

\subsection{Global response-direction similarity}
We concatenate the selected responses using the same probe order for both models,
\begin{align}
\mathbf{R}_{M_1}
&=\operatorname{concat}_{i\in\mathcal{I}_{12}}\mathbf{r}_{M_1,i},\\
\mathbf{R}_{M_2}
&=\operatorname{concat}_{i\in\mathcal{I}_{12}}\mathbf{r}_{M_2,i}.
\end{align}
The final score is
\begin{equation}
S(M_1,M_2)=\frac{1}{2}\left(1+
\frac{\mathbf{R}_{M_1}^{\top}\mathbf{R}_{M_2}}
{\|\mathbf{R}_{M_1}\|_2\|\mathbf{R}_{M_2}\|_2}
\right),
\label{eq:score}
\end{equation}
with zero cosine used only in the degenerate zero-norm case. Thus magnitude decides \emph{where} to compare, while the signed response-vector direction determines \emph{how similarly} the two models respond.

\subsection{Query and comparison cost}
Top-25 selection is performed after response extraction and therefore does not reduce the probability-evaluation budget. With $N=971$ and $T=13$, each checkpoint requires $(T+1)N=13{,}594$ probability-returning inputs. Once cached, checkpoints can be compared without further model inference. On a single NVIDIA RTX A5000 GPU with 24\~GB memory, cache construction takes 6.56 minutes per model, corresponding to about 2.08 GPU-hours for the 19 candidate checkpoints and 3.72 GPU-hours for all 34 benchmark checkpoints. Query reduction is not a contribution of the current method.

\section{Experimental Setup}
\subsection{Unified direct-parent benchmark}
We evaluate 34 open-weight checkpoints spanning Qwen, Mistral, Llama, Baichuan, DeepSeek, Falcon, CodeLlama, and related variants. The benchmark contains 22 suspect models with 22 documented direct-parent edges and a global registry of 19 candidate parents. For every suspect, all registry candidates are considered except exact self when applicable, yielding 411 suspect--candidate pairs. Candidate pools are not restricted by family, so siblings, ancestors, descendants, and cross-family checkpoints remain as hard negatives.

The parent labels were separately audited against public provenance evidence, independently of the BReF scoring procedure. All 22 edges have public support. We use three author-defined evidence-quality categories solely to organize the strength of public provenance documentation; they are not an external certification standard. Fifteen edges receive Grade A evidence, six Grade B evidence, and one conservative Grade C edge; the evidence criteria and edge-level audit are
documented in Appendix~\ref{app:benchmark}. All 34 checkpoints map to public repositories. Experimental repository revision hashes were not recorded at download time; we therefore report them as \texttt{NOT\_RECORDED} rather than inferring present-day repository heads. 

For relation-level summaries, we report 22 direct-parent (DP) pairs,
a 43-pair same-family non-parent subset (SF), and 337 different-family
(DF) pairs. An additional 9 non-parent pairs are retained in DP--All but
are not included in the relation-specific SF or DF summaries. The DP--SF
and DP--DF analyses use the corresponding subsets, while DP--All treats
all 389 non-parent registry candidates as negatives. Only the 22
documented direct parents count as correct retrieval targets.

\paragraph{Probe-pool construction.}
The 971 probes come from fixed text-only cleaning of a 1,014-item pool
constructed from QQP, MRPC, ANLI, and IFEval using predefined
dataset-specific A/B/C/D conversion templates. The verbal options are fixed
during probe construction rather than taken from the source datasets or
generated by the evaluated models. Appendix~\ref{app:probes} details the
dataset-specific conversion rules, source counts, text-quality filtering,
template-level deduplication, and label-metadata caveat.

\subsection{Metrics}
The primary cross-method metrics are Top-1 accuracy, Top-3 accuracy, mean reciprocal rank (MRR), and ROC-AUC, all of which are insensitive to the absolute numerical scale of a method's score. We additionally use a DP--DF mean-separation gap as a \emph{within-method} diagnostic. For BReF,
\begin{equation}
\mathrm{Gap}_{\mathrm{DP-DF}}
=
\mathbb{E}[S\mid\mathrm{DP}]
-
\mathbb{E}[S\mid\mathrm{DF}].
\end{equation}
where $S(M_1,M_2)=
\frac{1}{2}
\left(
1+\cos(\mathbf{R}_{M_1},\mathbf{R}_{M_2})
\right)\in[0,1]$ and larger values indicate greater behavioral similarity. The raw DP--DF gap is interpreted only within each method because the reproduced methods use different score semantics and numerical scales. Candidate ranking follows each method's native closeness direction. For ROC-AUC, distance-valued scores such as MET's MMD and LLMmap's cosine distance are sign-reversed so that larger values consistently indicate stronger parent compatibility. Cross-method comparisons therefore rely on Top-1, Top-3, MRR, and ROC-AUC rather than raw score gaps. For BReF we also report direct-parent versus all-non-parent AUC and the true-parent retrieval margin, defined for suspect $Q$ as
\begin{equation}
m(Q)
=
S(Q,P^\star)
-
\max_{\substack{P_j\in\mathcal{P}_Q\\P_j\neq P^\star}}
S(Q,P_j),
\label{eq:parent_margin}
\end{equation}
where $P^\star$ is the documented direct parent. Positive margin therefore means that the parent ranks first. Reported mean margins average $m(Q)$ over the relevant suspects. Because pair scores share suspects, uncertainty is estimated with suspect-level clustered resampling rather than treating all 411 pairs as IID.

\subsection{Baselines}
We reproduce REEF \citep{zhang2025reef}, LLMmap \citep{pasquini2025llmmap}, MET \citep{gao2025met}, and ZeroPrint \citep{shao2026zeroprint} on the same 22-case candidate protocol, preserving each method's native probing, representation, and score. We also construct an independent QA-Agreement baseline from the same 971 probes and clean13 perturbations using only A/B/C/D argmax agreement. Because the protocols use different signals and budgets, Table~\ref{tab:main_results} compares retrieval performance rather than equal-cost efficiency; exact reproduction details and score orientations are given in Appendix~\ref{app:baselines}.

\section{Results}

\subsection{Main direct-parent retrieval results}
\begin{figure*}[t]
\centering
\IfFileExists{\heatmapfigurepath}{%
    \includegraphics[width=0.96\textwidth]{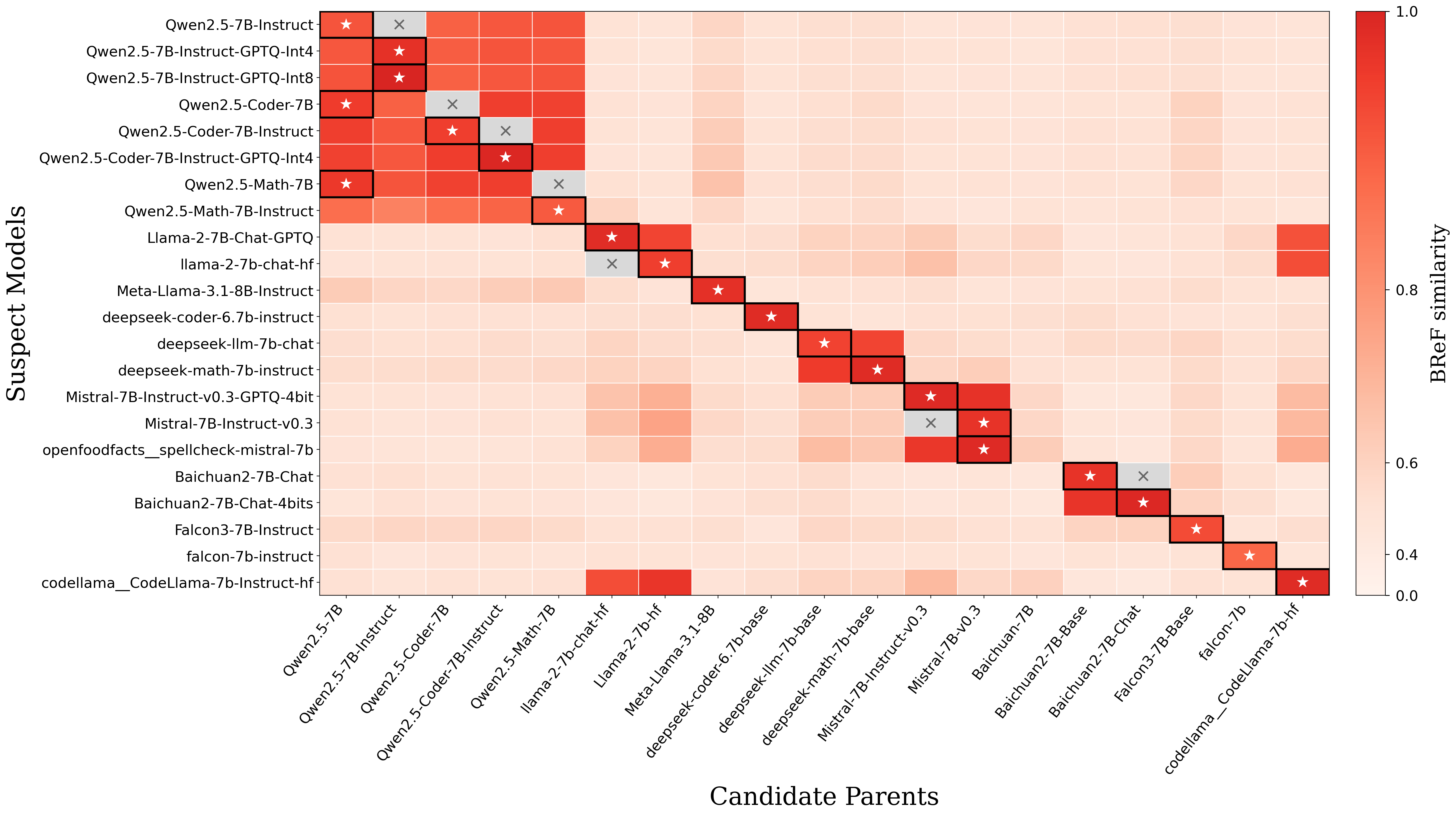}%
}{%
    \figplaceholder{2.2in}{\textbf{FINAL REPRODUCIBLE HEATMAP FILE NOT FOUND.}\\
    Expected path: \texttt{figures/bref\_exact\_parent\_heatmap.png}.}%
}
\caption{Exact-parent similarity heatmap on the unified direct-parent benchmark. Rows are suspect models and columns are candidate parents; darker red denotes higher BReF similarity. Black-outlined cells with a white star indicate the documented direct parent of each suspect model. Gray cells marked with $\times$ correspond to suspect--candidate pairs that were not evaluated in the benchmark and are treated as missing entries rather than zero-similarity pairs.}
\label{fig:exact_parent_heatmap}
\end{figure*}

BReF retrieves the documented parent at rank 1 for all 22 suspects
(Table~\ref{tab:main_results}), with Top-3 22/22, MRR 1.0000, and
DP--DF AUC 1.0000. The mean similarities for DP, SF, and DF pairs are
0.9668, 0.7873, and 0.5320, respectively, yielding a DP--DF gap of
0.4348. DP--SF AUC is 0.8969, DP--All AUC is 0.9832, and the mean
true-parent margin is 0.1105. REEF is the strongest reproduced prior
method (MRR 0.8171), while BReF achieves the highest Top-1, Top-3,
MRR, and DP--DF AUC. Figure~\ref{fig:exact_parent_heatmap}
visualizes the pair-score structure behind these retrieval results.
\begin{table*}[t]
\centering
\small
\setlength{\tabcolsep}{4.8pt}
\renewcommand{\arraystretch}{1.16}
\caption{Unified direct-parent retrieval performance. Raw score gaps are omitted because score scales differ across methods.}
\label{tab:main_results}
\begin{tabular*}{\textwidth}{@{\extracolsep{\fill}}lccrrrr@{}}
\toprule
Method & Signal & Top-1 $\uparrow$ & Top-3 $\uparrow$ & MRR $\uparrow$ & AUC$_{\mathrm{DP-DF}}$ $\uparrow$ & AUC$_{\mathrm{DP-All}}$ $\uparrow$ \\
\midrule
LLMmap & Text & 22.73 & 50.00 & 0.4104 & 0.7027 & 0.6938 \\
ZeroPrint & Text & 31.82 & 40.91 & 0.4234 & 0.7142 & 0.6933 \\
MET & Samples & 63.64 & 68.18 & 0.6962 & 0.7787 & 0.7752 \\
QA-Agreement & Hard labels & 63.64 & 72.73 & 0.7202 & 0.8567 & 0.8354 \\
REEF & Hidden repr. & 68.18 & 95.45 & 0.8171 & 0.9679 & 0.9507 \\
\textbf{BReF (ours)} & \textbf{Option prob.} & \textbf{100.00} & \textbf{100.00} & \textbf{1.0000} & \textbf{1.0000} & \textbf{0.9832} \\
\bottomrule
\end{tabular*}
\end{table*}

\subsection{Statistical reliability and paired significance}
Because the same checkpoints appear in many of the 411 pair scores, we do not treat pairs as independent. Suspect-level clustered bootstrap with 10,000 replicates gives DP--DF AUC $1.0000$ with 95\% CI $[1.0000,1.0000]$, DP--All AUC $0.9832$ with CI $[0.9736,0.9933]$, and a mean true-parent margin of 0.1105 with CI $[0.0526,0.1770]$. Additional clustered intervals are reported in Appendix~\ref{app:stats}.

For retrieval ranking, we bootstrap the 22 suspect cases as paired units. Relative to the strongest prior baseline REEF, BReF improves MRR by 0.1829 with 95\% CI $[0.0682,0.3052]$. The paired MRR improvements against MET, QA-Agreement, ZeroPrint, and LLMmap are also strictly positive. Exact paired McNemar tests on Top-1 correctness remain significant for all five comparisons after Holm correction (Table~\ref{tab:stats_main}).

A conservative sensitivity check that resamples the six model families as clusters keeps the $\Delta$MRR intervals and leave-one-family-out ranges strictly above zero for all five baselines; because only six family clusters are available, suspect-clustered inference remains primary (Appendix~\ref{app:stats}).

We also apply paired tests to the internal mechanism controls. Relative to the magnitude-only Top-25 control, BReF increases Top-1 from 14/22 to 22/22; the exact McNemar test gives $p=0.0078$ and the Holm-adjusted value is $p=0.0313$. Its paired MRR improvement is 0.245 with a 95\% bootstrap CI of $[0.108,0.396]$. The Top-1 differences against Query-only Top-25, All971 PLR, and Absolute-log$P$ do not reach significance under exact paired McNemar tests with 22 suspect cases. We therefore treat the evidence for response direction as statistically supported, while describing the incremental advantage of the pair-specific selector more cautiously.
\begin{table}[t]
\centering
\small
\setlength{\tabcolsep}{2.7pt}
\renewcommand{\arraystretch}{1.16}
\caption{Paired significance over 22 suspects. $\Delta$MRR is BReF minus comparator; Holm-$p$ is the corrected exact McNemar value.}
\label{tab:stats_main}
\begin{tabular*}{\columnwidth}{@{\extracolsep{\fill}}lrrr@{}}
\toprule
Comparator & $\Delta$MRR & 95\% CI & Holm-$p$ \\
\midrule
REEF & 0.1829 & [0.0682, 0.3052] & 0.023438 \\
MET & 0.3038 & [0.1428, 0.4770] & 0.023438 \\
QA-Agreement & 0.2798 & [0.1277, 0.4427] & 0.023438 \\
ZeroPrint & 0.5766 & [0.4066, 0.7355] & 0.000244 \\
LLMmap & 0.5896 & [0.4384, 0.7326] & 0.000076 \\
\bottomrule
\end{tabular*}
\end{table}

\subsection{Coarse separation versus exact-parent ranking}
Table~\ref{tab:mechanism_main} gives the central mechanism result. Static probability geometry is already informative: Absolute-log$P$ reaches DP--DF AUC 0.9706 but only 19/22 Top-1. Likewise, using all 971 PLR probes yields DP--DF AUC 0.9995 yet only 19/22 Top-1. Randomly selecting 25 probes also produces high coarse separation: over 50 seeds, Random-25 obtains DP--DF AUC $0.9822\pm0.0207$ but only $18.04\pm1.31$/22 Top-1. These results show that high cross-family AUC does not imply reliable exact-parent retrieval.

The same distinction remains within model families. For BReF, direct-parent versus same-family non-parent discrimination gives DP--SF AUC 0.8969, below its DP--DF AUC of 1.0000, confirming that closely related checkpoints form the harder regime. Exact retrieval nevertheless remains 22/22, showing that pooled relation discrimination and suspect-conditioned parent ranking measure different aspects of performance.

Pair-specific selection yields the best exact-parent ranking in the matched ablations. Query-only Top-25 reaches 21/22, Bottom-25 18/22, and All-971 19/22, while the pair-specific geometric Top-25 reaches 22/22. Its advantage is therefore best interpreted as concentrating the comparison on perturbation-sensitive probes that help resolve difficult sibling and derived checkpoints, rather than as maximizing pooled AUC. QA-Agreement reaches 14/22 Top-1 using the same 971 probes and 13 perturbations but only hard A/B/C/D decisions, showing that the probability-response geometry contributes information beyond answer agreement. 

\begin{table*}[t]
\centering
\small
\setlength{\tabcolsep}{3.5pt}
\renewcommand{\arraystretch}{1.16}
\caption{Matched representation and selector controls. Random-25 reports mean $\pm$ std over 50 seeds.}
\label{tab:mechanism_main}
\begin{tabular*}{\textwidth}{@{\extracolsep{\fill}}lrrrrr@{}}
\toprule
Method / control & Top-1 & Top-3 & MRR & DP--DF Gap & AUC$_{\mathrm{DP-DF}}$ \\
\midrule
\textbf{BReF Pair-Top25} & \textbf{22/22} & \textbf{22/22} & \textbf{1.000} & \textbf{0.4348} & \textbf{1.0000} \\
Query-only Top25 & 21/22 & 22/22 & 0.9697 & 0.4235 & 1.0000 \\
Random25 (50 seeds) & $18.04\pm1.31$ & $21.10\pm0.89$ & $0.8913\pm0.0428$ & $0.3109\pm0.0236$ & $0.9822\pm0.0207$ \\
Bottom25 & 18/22 & 21/22 & 0.8769 & 0.2598 & 0.9827 \\
All971 PLR & 19/22 & 22/22 & 0.9242 & 0.3188 & 0.9995 \\
Identity-$P$ & 17/22 & 21/22 & 0.8674 & 0.0968 & 0.9278 \\
Identity-log$P$ & 18/22 & 21/22 & 0.8879 & 0.0729 & 0.9527 \\
Absolute-$P$ & 18/22 & 21/22 & 0.8914 & 0.1029 & 0.9454 \\
Absolute-log$P$ & 19/22 & 21/22 & 0.9205 & 0.0804 & 0.9706 \\
QA agreement, all971 & 14/22 & 16/22 & 0.7202 & 0.2429 & 0.8567 \\
\bottomrule
\end{tabular*}
\end{table*}

\subsection{Energy-only controls and the role of response direction}

To isolate response magnitude from response direction, we construct controls from the response energy
\begin{equation}
E_M(i)=\|\mathbf{r}_{M,i}\|_2.
\end{equation}
The corresponding joint response strength for a model pair is
\begin{equation}
J_i(M_1,M_2)
=
\sqrt{E_{M_1}(i)E_{M_2}(i)}.
\end{equation}
For the Top-25 mean control, the final magnitude-only score is
\begin{equation}
S_{\mathrm{mean}}(M_1,M_2)
=
\frac{1}{25}
\sum_{i\in\mathcal{I}_{12}}
\sqrt{E_{M_1}(i)E_{M_2}(i)}.
\label{eq:energy_mean}
\end{equation}
Joint-strength controls perform poorly even when evaluated on the same pair-specific Top-25 probes used by BReF, indicating that strong responses alone are insufficient for reliable ranking.

We therefore also evaluate a magnitude-similarity control,
\begin{equation}
s_i^{E}
=
1-
\frac{
|E_{M_1}(i)-E_{M_2}(i)|
}{
E_{M_1}(i)+E_{M_2}(i)+\varepsilon
}.
\label{eq:energy_similarity}
\end{equation}
Using the same Top-25 selector, magnitude similarity reaches 14/22 Top-1 and MRR 0.7551, compared with 22/22 and MRR 1.0000 for BReF. The paired Top-1 difference remains significant after Holm correction ($p=0.0313$), and the MRR improvement is 0.2449 with a 95\% confidence interval of $[0.1080,0.3956]$. These results indicate that the signed direction of the perturbation-response vectors provides information beyond response magnitude alone. Table~\ref{tab:energy_main} summarizes the complete magnitude-only controls.
\begin{table}[t]
\centering
\scriptsize
\setlength{\tabcolsep}{1.2pt}
\renewcommand{\arraystretch}{1.16}
\caption{Magnitude-only controls that remove response-vector direction.}
\label{tab:energy_main}
\begin{tabular*}{\columnwidth}{@{\extracolsep{\fill}}lrrrr@{}}
\toprule
Control & Top-1 & Top-3 & MRR & \shortstack{AUC$_{\mathrm{DP-DF}}$} \\
\midrule
Joint strength, Top25 mean & 5/22 & 10/22 & 0.4108 & 0.7444 \\
Joint strength, Top25 min & 5/22 & 11/22 & 0.4287 & 0.7328 \\
Joint strength, all971 mean & 1/22 & 4/22 & 0.2037 & 0.5279 \\
Magnitude similarity, Top25 & 14/22 & 19/22 & 0.7551 & 0.8817 \\
Magnitude similarity, all971 & 10/22 & 11/22 & 0.5567 & 0.7769 \\
\textbf{BReF (direction)} & \textbf{22/22} & \textbf{22/22} & \textbf{1.000} & \textbf{1.0000} \\
\bottomrule
\end{tabular*}
\end{table}

\subsection{Probe-count and perturbation-coverage sensitivity}
\begin{figure*}[t]
\centering
\begin{minipage}{0.49\textwidth}
\centering
\includegraphics[width=\linewidth]{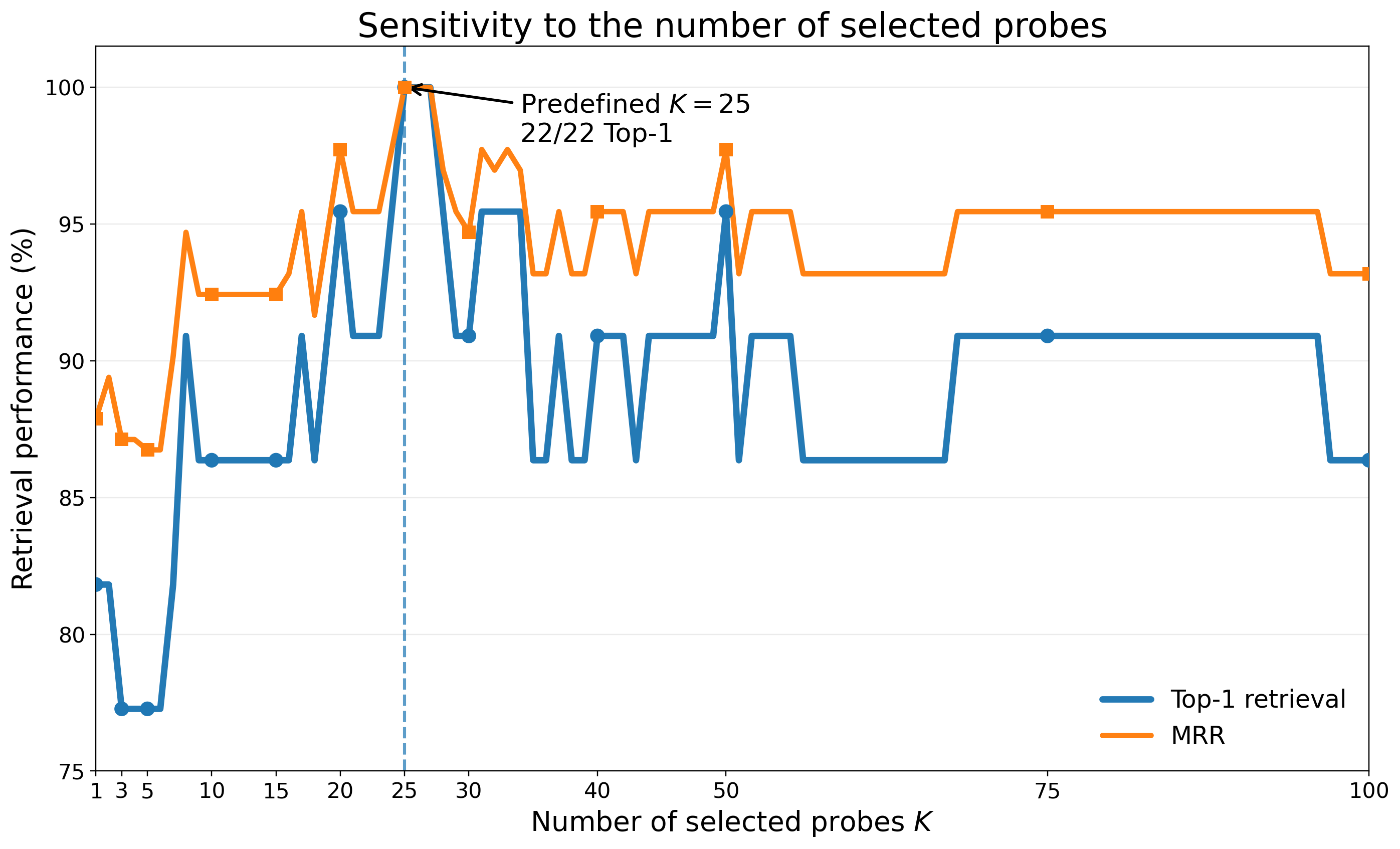}\vspace{-2pt}
{\small (a) Number of selected probes $K$.}
\end{minipage}\hfill
\begin{minipage}{0.49\textwidth}
\centering
\includegraphics[width=\linewidth]{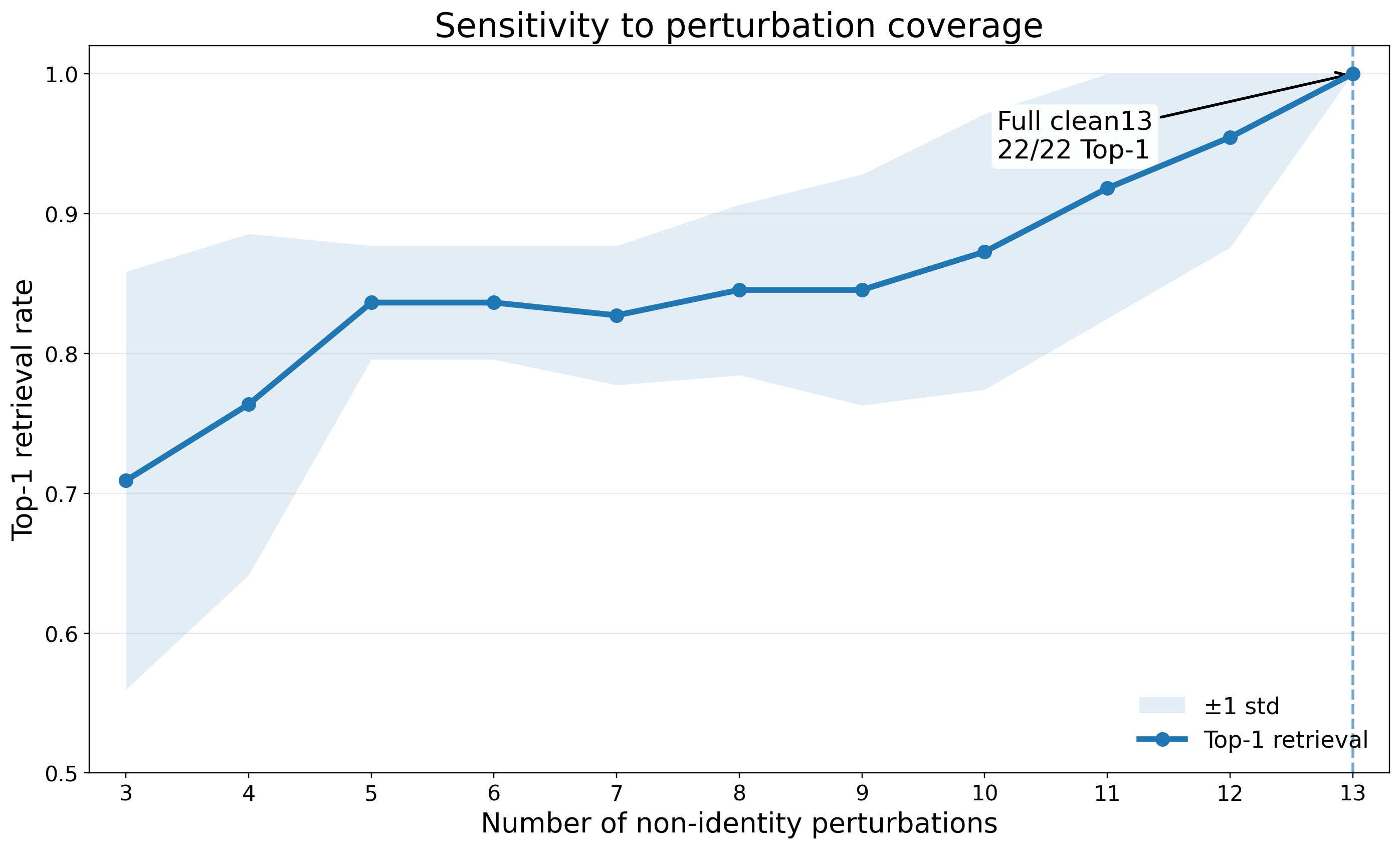}\vspace{-2pt}
{\small (b) Perturbation coverage.}
\end{minipage}
\caption{Sensitivity to selected-probe count and perturbation coverage. Panel (a) shows Top-1 retrieval and MRR as percentages; the dashed line marks $K=25$. Panel (b) shows mean Top-1 retrieval rate over five nested random perturbation orders, with shading denoting $\pm1$ standard deviation; the dashed line marks the full clean13 setting.}
\label{fig:sensitivity_curves}
\end{figure*}

Figure~\ref{fig:sensitivity_curves}(a) reports Top-1 retrieval and MRR as the number of selected probes varies. Exact-parent retrieval reaches 22/22 at $K=25$, 26, and 27; the predefined $K=25$ setting is the smallest evaluated operating point that attains perfect Top-1. Increasing $K$ further does not consistently improve suspect-conditioned ranking, and $K=100$ yields 19/22 Top-1. Selected reporting points are listed in Appendix~\ref{app:k_complete}.

Figure~\ref{fig:sensitivity_curves}(b) reports the mean Top-1 retrieval rate over five nested random perturbation orders, with the shaded region denoting $\pm1$ standard deviation. Performance is not strictly monotonic at every adjacent perturbation count, but the overall trend improves with coverage: the mean Top-1 rate rises from $15.6/22=70.9\%$ at $L=3$ to $21.0/22=95.5\%$ at $L=12$, and the full clean13 bank reaches $22/22=100\%$ with zero Top-1 variance across the five orders. The corresponding MRR statistics are reported in Appendix~\ref{app:ablation}.

\subsection{Semantic option permutation and calibration robustness}

Table~\ref{tab:robust_main} summarizes the semantic-option permutation
and asymmetric probability-calibration results.

\begin{table*}[!t]
\centering
\small
\setlength{\tabcolsep}{4.5pt}
\renewcommand{\arraystretch}{1.16}
\caption{Robustness to semantic option permutation and asymmetric probability
calibration. P0--P3 correspond to $(A,B,C,D)$, $(B,C,D,A)$,
$(D,C,B,A)$, and $(C,A,D,B)$, respectively.}
\label{tab:robust_main}
\begin{tabular*}{\textwidth}{@{\extracolsep{\fill}}lrrrlrrr@{}}
\toprule
\multicolumn{4}{c}{Semantic option permutation} &
\multicolumn{4}{c}{Asymmetric temperature $5\times5$ summary} \\
\cmidrule(lr){1-4}\cmidrule(lr){5-8}
Setting & Top-1 & Top-3 & MRR & Statistic & Min & Mean & Max \\
\midrule
P0 & 22/22 & 22/22 & 1.000 & Top-1 (/22) & 21 & 21.56 & 22 \\
P1 & 20/22 & 21/22 & 0.943 & Top-3 (/22) & 22 & 22.00 & 22 \\
P2 & 18/22 & 21/22 & 0.898 & MRR & 0.9773 & 0.9900 & 1.0000 \\
P3 & 20/22 & 22/22 & 0.947 & DP--DF AUC & 1.0000 & 1.0000 & 1.0000 \\
-- & -- & -- & -- & DP--DF Gap & 0.4327 & 0.4345 & 0.4355 \\
-- & -- & -- & -- & DP--SF AUC & 0.8964 & 0.8989 & 0.9054 \\
\bottomrule
\end{tabular*}
\end{table*}

Semantic permutation tests whether BReF relies on fixed A/B/C/D answer
positions rather than semantic response structure. P0 denotes the original
semantic option order $(A,B,C,D)$, while P1=$(B,C,D,A)$,
P2=$(D,C,B,A)$, and P3=$(C,A,D,B)$. The listed order specifies which
original semantic option is placed at each A/B/C/D label position. For each
permutation, we move the option contents to the corresponding label
positions, re-render every prompt, and evaluate the models again. The
returned A/B/C/D probabilities are then inverse-permuted back to the
original semantic coordinates before PLR construction. Thus, the displayed
answer positions change while the semantic coordinates compared by BReF
remain aligned across permutations.

P1, P2, and P3 retain 20/22, 18/22, and 20/22 Top-1, respectively;
their mean Top-1 is 87.90\%, and mean Top-3 is about 97.00\%.
All observed parent-rank degradations are confined to the Qwen family:
non-Qwen families retain perfect Top-1 across all four permutations, while
Qwen falls from 8/8 at P0 to 6/8, 4/8, and 6/8 under P1--P3.
These results show that BReF does not depend solely on a fixed
A/B/C/D-position shortcut, although exact within-family ranking remains
sensitive to semantic option placement in the most crowded Qwen regime.

For the calibration analysis, we apply temperature scaling directly to each
four-option probability vector before constructing the BReF representation.
For temperature $T$, the recalibrated probability of option $c$ is
\begin{equation}
\widetilde{p}_c(T)
=
\frac{p_c^{1/T}}
{\sum_{d\in\mathcal{Y}} p_d^{1/T}},
\; \forall c\in\mathcal{Y}.
\label{eq:temperature_calibration}
\end{equation}
The transformation is applied independently to every baseline and perturbed
probability vector before PLR construction. In the asymmetric experiment,
$T_q$ is applied to the suspect model and $T_c$ to the candidate model.
We then recompute the PLR responses, pair-specific Top-25 probe selection,
and global cosine similarity from the corresponding recalibrated
representations.
Across the asymmetric $5\times5$ calibration grid
$T_q,T_c\in\{0.70,0.85,1.00,1.15,1.30\}$,
Top-1 ranges from 21/22 to 22/22 (mean 21.56/22), all 25 settings retain
22/22 Top-3, and MRR ranges from 0.9773 to 1.0000. The DP--DF gap remains
in $[0.4327,0.4355]$ (mean 0.4345), DP--DF AUC is 1.0000 for all 25
settings, and DP--SF AUC ranges from 0.8964 to 0.9054 (mean 0.8989).
The $T_q=T_c=1$ setting recovers the original probability vectors.
These results indicate that moderate probability recalibration can perturb
some exact parent rankings while largely preserving provenance separation.

\subsection{Transformation-level coverage and hard cases}

The 22 verified parent edges span four transformation groups: chat/alignment
(3 edges), code/math/task fine-tuning (3), instruction tuning (10), and
quantization (6). BReF retrieves the documented parent at rank 1 for every
edge in all four groups, with MRR 1.000 throughout. The complete
transformation-level breakdown is reported in
Appendix~\ref{app:benchmark}. Under the main BReF setting, Qwen contributes
eight suspects and retains 8/8 Top-1 retrieval; family-clustered uncertainty
analysis is reported in Appendix~\ref{app:stats}.

\section{Protocol Audits and Reproducibility}
\label{sec:controls}
All 34 tokenizers represent unspaced A/B/C/D as distinct single tokens. Independent repeat runs of the two Falcon checkpoints produce bitwise-identical probability outputs, with maximum absolute difference 0 and Pearson correlation 1. The accompanying artifact and reproducibility materials are summarized in the Reproducibility Checklist.

\section{Discussion and Limitations}

\paragraph{Coarse separation and fine-grained ranking capture different properties.}
The control experiments show that strong pooled relation separation does not necessarily translate into equally strong candidate ranking. Static probabilities, all-probe PLR responses, and random probe subsets can preserve substantial relation-level discrimination while producing different ranking behavior. Among the tested mechanism controls, the clearest statistically supported difference is between magnitude-only comparison and directional PLR comparison, whereas the incremental effect of pair-specific probe selection is more modest with the current number of suspects.

\paragraph{Robustness is strong but not absolute.}
Semantic option permutations affect several Qwen-family rankings, whereas moderate probability rescaling causes smaller changes. BReF is therefore robust to these tested variations but not invariant to answer ordering or probability calibration.

\paragraph{Scope and reproducibility.}
The method requires access to complete four-option probabilities, or
sufficient option log-probabilities. The current benchmark does not address
ambiguous multi-parent merges.

\section{Conclusion}
BReF characterizes LLMs through perturbation-induced changes in four-option probability distributions, using response magnitude for probe selection and response direction for similarity. Across the evaluated benchmark and mechanism controls, perturbation-response geometry provides information beyond hard answer agreement and response magnitude alone, while same-family discrimination and answer-order sensitivity remain important challenges. These findings support perturbation-response behavior as a useful signal for fine-grained model provenance analysis.

\clearpage
\raggedbottom

\flushbottom
\clearpage
\appendix

\section{Probe Pool and Perturbation Disclosure}
\label{app:probes}
\paragraph{Probe-pool construction and source accounting.}
We first constructed a 1,014-item candidate pool from QQP, MRPC, ANLI,
and IFEval using fixed dataset-specific conversion templates. For QQP and
MRPC, each original sentence pair is rendered as \texttt{Text A} and
\texttt{Text B}. For ANLI, the premise--hypothesis pair is rendered as
\texttt{Premise} and \texttt{Claim}, while each IFEval instruction is
rendered under \texttt{Instruction}. Each converted item is paired with a
predefined set of verbal options associated with the A/B/C/D answer labels.
These verbal options are introduced during probe construction rather than
taken from the original datasets, and they are not generated by the
evaluated models; the option templates are fixed in the seed library before
model evaluation.The exact verbal-option templates are included in the released seed library.

Text-only quality control removes 42 items, and template-level
deduplication removes one additional near duplicate at a fixed lexical
Jaccard threshold of 0.92, yielding the final 971-probe pool. The complete
construction is performed independently of model performance on the current
evaluation benchmark.
\begin{table}[h]
\centering
\small
\setlength{\tabcolsep}{5pt}
\renewcommand{\arraystretch}{1.16}
\begin{tabular*}{\columnwidth}{@{\extracolsep{\fill}}lrrr@{}}
\toprule
Source & Initial & Removed & Final \\
\midrule
QQP & 370 & 21 & 349 \\
MRPC & 304 & 19 & 285 \\
ANLI & 220 & 3 & 217 \\
IFEval & 120 & 0 & 120 \\
\midrule
\textbf{Total} & \textbf{1,014} & \textbf{43} & \textbf{971} \\
\bottomrule
\end{tabular*}
\caption{Probe-pool source composition.}
\label{tab:probe_sources}
\end{table}

Neither filtering nor deduplication uses outputs, similarities, or
performance from any of the 34 evaluated checkpoints, nor do they use the
22 parent labels, family labels, or lineage labels.

All 971 final probes contain a complete A/B/C/D option structure. We do \emph{not} claim that gold-answer positions are balanced across A/B/C/D: among the 195 items for which an explicit label is recoverable in the current metadata, all are stored as label 0 (A), while the remaining 776 items do not share a uniformly stored gold-answer field. This does not enter the BReF score, which uses the complete four-option probability vector and its perturbation-induced changes rather than the gold-answer label. The semantic option-permutation experiment provides a direct end-to-end control for fixed answer-position dependence.

\paragraph{Exact clean13 perturbation bank.}
The baseline/original condition is stored separately from the 13 deterministic perturbation operators. The clean13 bank was fixed before the reported sensitivity analyses. Table~\ref{tab:clean13_exact} specifies the exact deterministic operations used in the implementation.

\begin{table*}[h]
\centering
\footnotesize
\setlength{\tabcolsep}{3.2pt}
\renewcommand{\arraystretch}{1.16}
\begin{tabular}{p{0.22\textwidth}p{0.73\textwidth}}
\toprule
Perturbation & Exact deterministic operation \\
\midrule
\mbox{Prefix space} & Prepend one ASCII space: \texttt{" " + x}. \\
\mbox{Suffix space} & Append one ASCII space: \texttt{x + " "}. \\
\mbox{Add instruction} & Prefix the exact string \texttt{Please answer carefully.\textbackslash n}. \\
\mbox{Add strict label} & Append the exact string \texttt{\textbackslash nReturn only A, B, C, or D.}. \\
\mbox{Option spacing} & Globally replace \texttt{A.}, \texttt{B.}, \texttt{C.}, and \texttt{D.} with \texttt{A .}, \texttt{B .}, \texttt{C .}, and \texttt{D .}, respectively. \\
\mbox{Compact spaces} & Replace every maximal run matching regex \texttt{[ \textbackslash t]+} with one ASCII space; newlines are unchanged. \\
\mbox{Double newline} & If the input contains \texttt{\textbackslash n}, replace every newline with two newlines; otherwise append one newline. \\
\mbox{Answer $\rightarrow$ label} & Literal replacements \texttt{Answer:}$\rightarrow$\texttt{Label:} and \texttt{answer:}$\rightarrow$\texttt{label:}. \\
\mbox{Label $\rightarrow$ answer} & Literal replacements \texttt{Label:}$\rightarrow$\texttt{Answer:} and \texttt{label:}$\rightarrow$\texttt{answer:}. \\
\mbox{Which $\rightarrow$ choose} & Case-insensitive regex replacements: word-bounded \texttt{Which option}$\rightarrow$\texttt{Choose the option}; \texttt{Which choice}$\rightarrow$\texttt{Choose the choice}; \texttt{best fits}$\rightarrow$\texttt{matches best}. \\
\mbox{Choose $\rightarrow$ select} & Case-insensitive word-bounded \texttt{Choose}$\rightarrow$\texttt{Select}; and \texttt{Return only}$\rightarrow$\texttt{Output only}. \\
\mbox{Final-label phrase} & If the input contains literal substring \texttt{Final label} or \texttt{final label}, leave it unchanged; otherwise append \texttt{\textbackslash nFinal label:}. \\
\mbox{Option-style parentheses} & Globally replace \texttt{A.}, \texttt{B.}, \texttt{C.}, and \texttt{D.} with \texttt{(A)}, \texttt{(B)}, \texttt{(C)}, and \texttt{(D)}, respectively. \\
\bottomrule
\end{tabular}
\caption{Exact deterministic clean13 perturbations used by BReF.}
\label{tab:clean13_exact}
\end{table*}

The executable transformation functions are released with the artifact.

\paragraph{Probability extraction.}
The protocol uses unspaced A/B/C/D answer labels. All 34 evaluated tokenizers map these four labels to distinct single tokens. The scoring procedure requires only the four option probabilities under the common prompt protocol; token IDs are needed only for the local extraction audit.

\section{Unified Direct-Parent Benchmark}
\label{app:benchmark}

The final benchmark contains 34 checkpoints, 22 suspects, 19 global
candidate parents, 22 documented direct-parent edges, and 411 valid
suspect--candidate pairs. Exact self-candidates are removed where
applicable, while same-family hard negatives are retained. The
transformation composition of the 22 parent edges is summarized in
Table~\ref{tab:transform_main}, and Table~\ref{tab:model_registry}
enumerates all 34 evaluated checkpoints with links to their public
Hugging Face repositories.

\begin{table}[h]
\centering
\footnotesize
\setlength{\tabcolsep}{2.6pt}
\renewcommand{\arraystretch}{1.16}
\caption{Direct-parent retrieval by transformation type.}
\label{tab:transform_main}
\begin{tabular*}{\columnwidth}{@{\extracolsep{\fill}}lrrr@{}}
\toprule
Transformation type & Edges & Top-1 & MRR \\
\midrule
Chat / alignment & 3 & 3/3 & 1.000 \\
Code / math / task FT & 3 & 3/3 & 1.000 \\
Instruction tuning & 10 & 10/10 & 1.000 \\
Quantization & 6 & 6/6 & 1.000 \\
\bottomrule
\end{tabular*}
\end{table}

\begin{table*}[t]
\centering
\scriptsize
\setlength{\tabcolsep}{2.4pt}
\renewcommand{\arraystretch}{1.12}
\begin{tabular}{r >{\raggedright\arraybackslash}p{0.42\textwidth} r >{\raggedright\arraybackslash}p{0.42\textwidth}}
\toprule
\# & Public checkpoint repository & \# & Public checkpoint repository \\
\midrule
1 & \href{https://huggingface.co/Qwen/Qwen2.5-7B}{\nolinkurl{Qwen/Qwen2.5-7B}} & 18 & \href{https://huggingface.co/deepseek-ai/deepseek-coder-6.7b-base}{\nolinkurl{deepseek-ai/deepseek-coder-6.7b-base}} \\
2 & \href{https://huggingface.co/Qwen/Qwen2.5-7B-Instruct}{\nolinkurl{Qwen/Qwen2.5-7B-Instruct}} & 19 & \href{https://huggingface.co/deepseek-ai/deepseek-coder-6.7b-instruct}{\nolinkurl{deepseek-ai/deepseek-coder-6.7b-instruct}} \\
3 & \href{https://huggingface.co/Qwen/Qwen2.5-7B-Instruct-GPTQ-Int4}{\nolinkurl{Qwen/Qwen2.5-7B-Instruct-GPTQ-Int4}} & 20 & \href{https://huggingface.co/deepseek-ai/deepseek-llm-7b-base}{\nolinkurl{deepseek-ai/deepseek-llm-7b-base}} \\
4 & \href{https://huggingface.co/Qwen/Qwen2.5-7B-Instruct-GPTQ-Int8}{\nolinkurl{Qwen/Qwen2.5-7B-Instruct-GPTQ-Int8}} & 21 & \href{https://huggingface.co/deepseek-ai/deepseek-llm-7b-chat}{\nolinkurl{deepseek-ai/deepseek-llm-7b-chat}} \\
5 & \href{https://huggingface.co/Qwen/Qwen2.5-Coder-7B}{\nolinkurl{Qwen/Qwen2.5-Coder-7B}} & 22 & \href{https://huggingface.co/deepseek-ai/deepseek-math-7b-base}{\nolinkurl{deepseek-ai/deepseek-math-7b-base}} \\
6 & \href{https://huggingface.co/Qwen/Qwen2.5-Coder-7B-Instruct}{\nolinkurl{Qwen/Qwen2.5-Coder-7B-Instruct}} & 23 & \href{https://huggingface.co/deepseek-ai/deepseek-math-7b-instruct}{\nolinkurl{deepseek-ai/deepseek-math-7b-instruct}} \\
7 & \href{https://huggingface.co/Qwen/Qwen2.5-Coder-7B-Instruct-GPTQ-Int4}{\nolinkurl{Qwen/Qwen2.5-Coder-7B-Instruct-GPTQ-Int4}} & 24 & \href{https://huggingface.co/meta-llama/Llama-2-7b-chat-hf}{\nolinkurl{meta-llama/Llama-2-7b-chat-hf}} \\
8 & \href{https://huggingface.co/Qwen/Qwen2.5-Math-7B}{\nolinkurl{Qwen/Qwen2.5-Math-7B}} & 25 & \href{https://huggingface.co/meta-llama/Llama-2-7b-hf}{\nolinkurl{meta-llama/Llama-2-7b-hf}} \\
9 & \href{https://huggingface.co/Qwen/Qwen2.5-Math-7B-Instruct}{\nolinkurl{Qwen/Qwen2.5-Math-7B-Instruct}} & 26 & \href{https://huggingface.co/meta-llama/Llama-3.1-8B}{\nolinkurl{meta-llama/Llama-3.1-8B}} \\
10 & \href{https://huggingface.co/RedHatAI/Mistral-7B-Instruct-v0.3-GPTQ-4bit}{\nolinkurl{RedHatAI/Mistral-7B-Instruct-v0.3-GPTQ-4bit}} & 27 & \href{https://huggingface.co/meta-llama/Llama-3.1-8B-Instruct}{\nolinkurl{meta-llama/Llama-3.1-8B-Instruct}} \\
11 & \href{https://huggingface.co/TheBloke/Llama-2-7B-Chat-GPTQ}{\nolinkurl{TheBloke/Llama-2-7B-Chat-GPTQ}} & 28 & \href{https://huggingface.co/mistralai/Mistral-7B-Instruct-v0.3}{\nolinkurl{mistralai/Mistral-7B-Instruct-v0.3}} \\
12 & \href{https://huggingface.co/baichuan-inc/Baichuan-7B}{\nolinkurl{baichuan-inc/Baichuan-7B}} & 29 & \href{https://huggingface.co/mistralai/Mistral-7B-v0.3}{\nolinkurl{mistralai/Mistral-7B-v0.3}} \\
13 & \href{https://huggingface.co/baichuan-inc/Baichuan2-7B-Base}{\nolinkurl{baichuan-inc/Baichuan2-7B-Base}} & 30 & \href{https://huggingface.co/openfoodfacts/spellcheck-mistral-7b}{\nolinkurl{openfoodfacts/spellcheck-mistral-7b}} \\
14 & \href{https://huggingface.co/baichuan-inc/Baichuan2-7B-Chat}{\nolinkurl{baichuan-inc/Baichuan2-7B-Chat}} & 31 & \href{https://huggingface.co/tiiuae/Falcon3-7B-Base}{\nolinkurl{tiiuae/Falcon3-7B-Base}} \\
15 & \href{https://huggingface.co/baichuan-inc/Baichuan2-7B-Chat-4bits}{\nolinkurl{baichuan-inc/Baichuan2-7B-Chat-4bits}} & 32 & \href{https://huggingface.co/tiiuae/Falcon3-7B-Instruct}{\nolinkurl{tiiuae/Falcon3-7B-Instruct}} \\
16 & \href{https://huggingface.co/codellama/CodeLlama-7b-Instruct-hf}{\nolinkurl{codellama/CodeLlama-7b-Instruct-hf}} & 33 & \href{https://huggingface.co/tiiuae/falcon-7b}{\nolinkurl{tiiuae/falcon-7b}} \\
17 & \href{https://huggingface.co/codellama/CodeLlama-7b-hf}{\nolinkurl{codellama/CodeLlama-7b-hf}} & 34 & \href{https://huggingface.co/tiiuae/falcon-7b-instruct}{\nolinkurl{tiiuae/falcon-7b-instruct}} \\
\bottomrule
\end{tabular}
\caption{Public checkpoint registry for all 34 evaluated models. Each displayed repository identifier is a clickable link to the exact public Hugging Face model page corresponding to the checkpoint used in the benchmark. For the Meta Llama 3.1 models, we use Hugging Face's current canonical repository paths. Experiment-time repository revision hashes were not recorded, so the links identify the public repositories rather than claiming that their present-day HEAD revisions are byte-identical to the acquired checkpoints.}
\label{tab:model_registry}
\end{table*}

\paragraph{Provenance audit.}
All 22 edges have public evidence: 15 Grade A, six Grade B, and one Grade C. These are author-defined evidence-quality categories used only to organize the strength of public provenance documentation; they are not an external certification standard. Grade A denotes structured evidence such as a model tree, explicit \texttt{base\_model} metadata, or quantization lineage; Grade B denotes explicit official natural-language derivation statements; the single Baichuan2 edge is conservatively Grade C because the official release identifies corresponding base and aligned/chat models without equally strong machine-readable parent metadata. All 34 checkpoints map to the public repositories listed in Table~\ref{tab:model_registry}.

\begin{table}[h]
\centering
\small
\setlength{\tabcolsep}{3.2pt}
\renewcommand{\arraystretch}{1.16}
\begin{tabular*}{\columnwidth}{@{\extracolsep{\fill}}lrrrr@{}}
\toprule
\shortstack[l]{Evidence\\subset} & Edges & Top-1 & Top-3 & MRR \\
\midrule
Grade A & 15 & 15/15 & 15/15 & 1.000 \\
Grade A+B & 21 & 21/21 & 21/21 & 1.000 \\
Grade A+B+C & 22 & 22/22 & 22/22 & 1.000 \\
\bottomrule
\end{tabular*}
\caption{Sensitivity to provenance evidence quality.}
\label{tab:provenance_quality}
\end{table}

These results show that the perfect exact-parent retrieval result is not driven by the single lower-confidence provenance edge.

\section{Baseline Reproduction Details}
\label{app:baselines}

All reproduced baselines are evaluated on the same 22 suspects and global candidate-parent registry, with exact-self candidates removed and same-family hard negatives retained. The original fingerprinting signal of each method is preserved; only the final score is used to rank the common candidate registry. Table~\ref{tab:baseline_protocols} summarizes the resulting protocols. The budgets are reported for transparency and are not intended to imply equal computational cost.

\begin{table*}[t]
\centering
\footnotesize
\setlength{\tabcolsep}{3pt}
\renewcommand{\arraystretch}{1.14}
\begin{tabularx}{\textwidth}{@{}l >{\raggedright\arraybackslash}X >{\raggedright\arraybackslash}X l l@{}}
\toprule
Method & Per-checkpoint evaluation & Representation & Pair score & Parent ranking \\
\midrule
REEF & 200 TruthfulQA forwards & layer-18 last-token activations & linear CKA & larger first \\
LLMmap & 8 fixed generated queries & released 384-D fingerprint & cosine distance & smaller first \\
MET & 25 prompts $\times$ 10 samples & Unicode/Hamming output distribution & unbiased MMD & smaller first \\
ZeroPrint & 10 prompts $\times$ 20 samples & 589,824-D ZO fingerprint & Pearson correlation & larger first \\
QA-Agreement & $971\times13$ hard decisions & A/B/C/D argmax agreement & agreement fraction & larger first \\
\bottomrule
\end{tabularx}
\caption{Reproduction protocols used in the unified direct-parent benchmark. Query/sample counts describe the reproduced protocol for one checkpoint; they are not equal-cost budgets. For ROC-AUC, distance-valued scores are sign-reversed only to align score direction.}
\label{tab:baseline_protocols}
\end{table*}

\subsection{REEF}
We use the repository-local TruthfulQA CSV and take the first 200 statements in their original order, without shuffling, perturbation, generation, or repeated sampling. Each statement is forwarded once through each checkpoint. We hook transformer block \emph{index} 18 (Python zero-based indexing) and extract the full last-token vector $h[0,-1,:]$, giving an activation matrix $X\in\mathbb{R}^{200\times d}$ for each model.

Following the validated REEF preprocessing convention, features are centered across the 200 queries and divided by \texttt{torch.std(X, dim=0)} (PyTorch's default correction of 1); no numerical epsilon is added, and all 34 evaluated models have zero zero-variance feature dimensions. We then form the linear Gram matrix $K=XX^\top$ and center it with $H=I-\mathbf{1}\mathbf{1}^\top/n$. Writing $K_c=HKH$ and $L_c=HLH$, the score is
\begin{equation}
\mathrm{CKA}(K,L)=
\frac{\langle K_c,L_c\rangle_F}
{\|K_c\|_F\,\|L_c\|_F}.
\end{equation}
For every suspect--candidate pair, linear CKA is used directly as the ranking score and candidates are sorted in descending order, with candidate name as a deterministic tie-break. No classifier, threshold, BReF probe, BReF probability feature, or edge-specific calibration is introduced. The reproduced Qwen2.5-7B/Qwen2.5-7B-Instruct layer-index-18 CKA is 0.815955638885498, exactly matching the validated reference value used in our regression audit.

Compatibility handling is limited to software/checkpoint execution for legacy checkpoint and cache APIs; it does not change checkpoint weights, the 200 queries, layer index, last-token representation, or CKA definition. The protocol uses 200 forwards per checkpoint (6,800 for 34 models) and scores 411 cached activation pairs offline. REEF obtains Top-1 15/22, Top-3 21/22, MRR 0.8171, DP--DF AUC 0.9679, and DP--All AUC 0.9507.

\subsection{LLMmap}
LLMmap is reproduced with eight fixed queries per checkpoint. Responses are generated greedily with at most 256 new tokens and truncated to 650 characters following the released implementation. Each query and its response are separately embedded with \texttt{multilingual-E5-large-instruct} into 1,024-dimensional vectors; concatenating the query and response representations gives eight 2,048-dimensional traces. These traces are passed through the released open-set encoder to obtain a 384-dimensional model fingerprint.

The released cosine \emph{distance} is retained as the raw pair score. For each suspect, all candidate parents are therefore ranked in ascending distance. When computing ROC-AUC, we use the sign-reversed distance only to orient the statistic so that larger values mean stronger parent compatibility; this monotonic transformation does not change ranking. Once candidate fingerprints are precomputed, attribution of a suspect requires eight generated-response queries. On the common registry, LLMmap obtains Top-1 5/22, Top-3 11/22, MRR 0.4104, DP--DF AUC 0.7027, and DP--All AUC 0.6938.

\subsection{Model Equality Testing}
MET is reproduced with 25 multilingual Wikipedia prompts and 10 sampled completions per prompt, giving 250 completions per checkpoint. Generation uses temperature 1.0, top-$p=1.0$, top-$k=0$, and at most 50 new tokens. Outputs are represented as Unicode-character sequences with length $L=1000$ and right-padding value $-1$. We use the normalized prompt-aware Hamming kernel and the unbiased empirical maximum mean discrepancy (MMD), together with 100 permutation samples as in the reproduced testing protocol.

For direct-parent retrieval, the raw MMD statistic is used as a distance and candidate parents are ranked in ascending MMD. ROC-AUC is computed from $-\mathrm{MMD}$ so that larger values consistently denote stronger parent compatibility. No rescaling is used to compare its raw magnitude with BReF. MET obtains Top-1 14/22, Top-3 15/22, MRR 0.6962, DP--DF AUC 0.7787, and DP--All AUC 0.7752. Because the native MMD scale is not commensurate with BReF's normalized similarity, we do not report a cross-method raw DP--DF gap.

\subsection{ZeroPrint}
We reproduce the official ZeroPrint implementation at frozen commit \nolinkurl{16a02aa4cfd5693ecfa757e9d2832b7e9babada0}. The query reservoir is \texttt{openai\_humaneval}. Following the reproduced preprocessing, each selected HumanEval prompt is truncated to its first 20 words, prefixed with \texttt{Complete the following code:}, and filtered to a minimum final length of 15 words using the original reservoir-sampling/shuffling procedure. Two original queries are retained. For each query, four perturbations are generated by GloVe word substitution using \texttt{glove-wiki-gigaword-100}: three words are replaced per perturbation from the top 10 candidate neighbors. This yields $2\times(1+4)=10$ unique prompts.

Each prompt is sampled 20 times, for 200 generations per checkpoint and 6,800 generations across the 34 checkpoints. Generation uses maximum input length 512, maximum 512 new tokens, temperature 0.7, top-$p=0.9$, top-$k=50$, and sampling enabled. Base models receive the raw prompt; instruction-tuned models use the reproduced instruct wrapper and the tokenizer chat template when available. Generated outputs are encoded with \texttt{sentence-transformers/all-mpnet-base-v2} into 768-dimensional vectors and averaged over the 20 stochastic generations for each prompt.

For each original query and its four perturbations, ZeroPrint forms changes in input-sentence embeddings $X$ and changes in mean output embeddings $Y$, estimates a zeroth-order Jacobian with ridge regression ($\alpha=0.001$), and flattens the Jacobian into a query-level fingerprint. The two query-level fingerprints are averaged, producing a 589,824-dimensional mean fingerprint. Pair similarity is Pearson correlation; for descriptive reporting it can be mapped to $[0,1]$ by $(\rho+1)/2$, without changing rankings.

GPU batch-size reductions and compatibility handling for legacy checkpoint/runtime APIs change only execution and not the query bank, number of generations, decoding, output embedding, Jacobian estimation, aggregation, or similarity definition. All 34 checkpoints produce valid fingerprints. Under the common benchmark, ZeroPrint obtains Top-1 7/22, Top-3 9/22, MRR 0.4234, DP--DF AUC 0.7142, and DP--All AUC 0.6933. Its mean similarities are 0.6325 for direct-parent and 0.5510 for different-family pairs, giving a within-method DP--DF gap of 0.0815. This gap is interpreted only within ZeroPrint's own score scale.

\subsection{QA-Agreement}
The independent QA-Agreement baseline uses the same 971 probes and the same 13 perturbation conditions as BReF, but deliberately discards probability magnitudes and all perturbation-response geometry. For model $M$, probe $i$, and perturbation $t$, we define the hard decision as
\begin{equation}
a_M(i,t)
=
\arg\max_{c\in\mathcal{Y}} p_{M,i,t,c}.
\end{equation}
The QA-Agreement score between two models is
\begin{equation}
\begin{aligned}
S_{\mathrm{QA}}(M_1,M_2)
&=\frac{1}{971\times13}
\sum_{i=1}^{971}\sum_{t=1}^{13}\\
&\quad \mathbf{1}\!\left[a_{M_1}(i,t)=a_{M_2}(i,t)\right].
\end{aligned}
\end{equation}
Thus the denominator is 12,623 hard decisions. The primary QA score uses only the 13 perturbed conditions and does not use PLR or the BReF pair-specific selector.

The QA-Agreement baseline using all 971 probes and 13 perturbations obtains Top-1 14/22, Top-3 16/22, MRR 0.720231, DP--DF AUC 0.856690, DP--All AUC 0.835359, and DP--SF AUC 0.713002. Its DP and DF mean similarities are 0.650627 and 0.407699, giving a DP--DF gap of 0.242928; the mean parent margin is 0.027587. Against BReF, eight suspects are correct only for BReF and none only for QA-Agreement, giving an uncorrected exact two-sided McNemar $p=0.0078125$. The paired MRR difference is 0.279769 with a 95\% bootstrap CI of $[0.127652,0.442687]$; the Holm-adjusted Top-1 comparison is reported in Table~\ref{tab:stats_main}.

\section{Ablations and Method Definition Audits}
\label{app:ablation}
\subsection{K-sensitivity details}
\label{app:k_complete}
Figure~\ref{fig:sensitivity_curves}(a) visualizes the ranking trend as the number of selected perturbation-sensitive probes varies. All other BReF components are held fixed. Table~\ref{tab:k_ranking} lists the principal reporting points, including the neighboring perfect-retrieval settings $K=26$ and $K=27$.

\begin{table}[h]
\centering
\small
\setlength{\tabcolsep}{4pt}
\renewcommand{\arraystretch}{1.16}
\begin{tabular*}{\columnwidth}{@{\extracolsep{\fill}}crr@{}}
\toprule
$K$ & Top-1 & MRR \\
\midrule
1 & 18/22 & 0.8788 \\
3 & 17/22 & 0.8712 \\
5 & 17/22 & 0.8674 \\
10 & 19/22 & 0.9242 \\
15 & 19/22 & 0.9242 \\
20 & 21/22 & 0.9773 \\
\textbf{25} & \textbf{22/22} & \textbf{1.0000} \\
26 & 22/22 & 1.0000 \\
27 & 22/22 & 1.0000 \\
30 & 20/22 & 0.9470 \\
40 & 20/22 & 0.9545 \\
50 & 21/22 & 0.9773 \\
75 & 20/22 & 0.9545 \\
100 & 19/22 & 0.9318 \\
\bottomrule
\end{tabular*}
\caption{Selected reporting points from the $K$-sensitivity analysis.}
\label{tab:k_ranking}
\end{table}

\subsection{Perturbation-count stability}
Figure~\ref{fig:sensitivity_curves}(b) plots Top-1 as a retrieval rate; the numerical values below are the equivalent correct-suspect counts out of 22 together with MRR. Over five nested random orders, Top-1 / MRR (mean $\pm$ std) are: $15.6\pm3.29$ / $0.8081\pm0.1055$ at $L=3$; $16.8\pm2.68$ / $0.8478\pm0.0843$ at $L=4$; $18.4\pm0.89$ / $0.8977\pm0.0257$ at $L=5$; $18.4\pm0.89$ / $0.8992\pm0.0248$ at $L=6$; $18.2\pm1.10$ / $0.8962\pm0.0280$ at $L=7$; $18.6\pm1.34$ / $0.9091\pm0.0377$ at $L=8$; $18.6\pm1.82$ / $0.9098\pm0.0498$ at $L=9$; $19.2\pm2.17$ / $0.9250\pm0.0594$ at $L=10$; $20.2\pm2.05$ / $0.9515\pm0.0568$ at $L=11$; $21.0\pm1.73$ / $0.9735\pm0.0476$ at $L=12$; and $22.0\pm0$ / $1.0000\pm0$ at $L=13$.

\subsection{Selector and static controls}
\begin{table*}[h]
\centering
\small
\setlength{\tabcolsep}{3.2pt}
\renewcommand{\arraystretch}{1.16}
\begin{tabular*}{\textwidth}{@{\extracolsep{\fill}}lrrrrr@{}}
\toprule
Control & Top-1 & Top-3 & MRR & DP--DF Gap & AUC$_{\mathrm{DP-DF}}$ \\
\midrule
BReF Pair-Top25 & 22/22 & 22/22 & 1.0000 & 0.4348 & 1.0000 \\
Query-only Top25 & 21/22 & 22/22 & 0.9697 & 0.4235 & 1.0000 \\
Random25 (50 seeds) & $18.04\pm1.31$ & $21.10\pm0.89$ & $0.8913\pm0.0428$ & $0.3109\pm0.0236$ & $0.9822\pm0.0207$ \\
Bottom25 & 18/22 & 21/22 & 0.8769 & 0.2598 & 0.9827 \\
All971 PLR & 19/22 & 22/22 & 0.9242 & 0.3188 & 0.9995 \\
Identity-$P$ & 17/22 & 21/22 & 0.8674 & 0.0968 & 0.9278 \\
Identity-log$P$ & 18/22 & 21/22 & 0.8879 & 0.0729 & 0.9527 \\
Absolute-$P$ & 18/22 & 21/22 & 0.8914 & 0.1029 & 0.9454 \\
Absolute-log$P$ & 19/22 & 21/22 & 0.9205 & 0.0804 & 0.9706 \\
\bottomrule
\end{tabular*}
\caption{Complete matched selector and static-probability controls.}
\end{table*}

For the selector variants, DP--All AUC is 0.9832 for Pair-Top25, 0.9783 for Query-only Top25, 0.9666 for Bottom25, and 0.9864 for All971 PLR.

\paragraph{Random25 variability.}
Across 50 fixed seeds, Random25 gives Top-1 $18.04\pm1.31$/22 (range 15--21), Top-3 $21.10\pm0.89$/22 (range 19--22), DP--All AUC $0.9683\pm0.0196$, and DP--SF AUC $0.8970\pm0.0222$. This reinforces the distinction between strong coarse separation and unstable exact-parent ranking.

\paragraph{Energy-only controls.}
We evaluate three joint-strength controls. The Top-25 mean control selects the pair-specific Top-25 probes by $\sqrt{E_{M_1}(i)E_{M_2}(i)}$ and uses the mean selected joint strength as the score; it obtains 5/22 Top-1, 10/22 Top-3, MRR 0.4108, and DP--DF AUC 0.7444. The Top-25 minimum control uses the minimum selected joint strength and obtains 5/22 Top-1, 11/22 Top-3, MRR 0.4287, and DP--DF AUC 0.7328. The all-probe mean control averages joint strength across all 971 probes and obtains 1/22 Top-1, 4/22 Top-3, MRR 0.2037, and DP--DF AUC 0.5279.

A magnitude-similarity control instead compares relative response magnitudes using Eq.~(\ref{eq:energy_similarity}). With the same pair-specific Top-25 selector it obtains 14/22 Top-1, 19/22 Top-3, MRR 0.7551, DP--DF gap 0.1747, and DP--DF AUC 0.8817; using all 971 probes obtains 10/22 Top-1, 11/22 Top-3, MRR 0.5567, DP--DF gap 0.1619, and DP--DF AUC 0.7769. These controls distinguish three different quantities: joint response strength identifies probes on which both models react strongly, magnitude similarity measures whether they react by similar amounts, and BReF's cosine compares the signed direction of the full perturbation-response vectors.

\section{Robustness Experiments}
\label{app:robustness}
\subsection{Semantic option permutation}

We use the four semantic permutations defined in the main text.

\begin{table}[h]
\centering
\small
\setlength{\tabcolsep}{3pt}
\renewcommand{\arraystretch}{1.16}
\begin{tabular*}{\columnwidth}{@{\extracolsep{\fill}}lrrrr@{}}
\toprule
Permutation & Top-1 & Top-3 & MRR & Qwen Top-1 \\
\midrule
P0 & 22/22 & 22/22 & 1.000 & 8/8 \\
P1 & 20/22 & 21/22 & 0.943 & 6/8 \\
P2 & 18/22 & 21/22 & 0.898 & 4/8 \\
P3 & 20/22 & 22/22 & 0.947 & 6/8 \\
\bottomrule
\end{tabular*}
\caption{Fresh semantic-option permutations. All observed Top-1 degradations are confined to the Qwen family.}
\end{table}
The principal rank changes involve Qwen2.5-7B-Instruct, Qwen2.5-Coder-7B, Qwen2.5-Math-7B, and Qwen2.5-Math-7B-Instruct. Other evaluated families retain perfect Top-1 under P0--P3.

\subsection{Probability calibration / temperature robustness}
Using the recalibration in Eq.~(\ref{eq:temperature_calibration}), across the
asymmetric $5\times5$ grid, Top-1 ranges 21--22/22 (mean 21.56), all
settings retain 22/22 Top-3, MRR ranges 0.977273--1.000000, DP--DF gap
ranges 0.432705--0.435491 (mean 0.434465), DP--DF AUC is 1.000000 for all
25 settings, and DP--SF AUC ranges 0.896406--0.905391
(mean 0.898879). The $T_q=T_c=1$ setting matches the main benchmark.

\section{Statistical Tests}
\label{app:stats}

\begin{table}[h]
\centering
\small
\setlength{\tabcolsep}{4pt}
\renewcommand{\arraystretch}{1.16}
\begin{tabular*}{\columnwidth}{@{\extracolsep{\fill}}lr@{}}
\toprule
Final BReF metric & Point estimate \\
\midrule
Top-1 & 22/22 \\
Top-3 & 22/22 \\
MRR & 1.000000 \\
DP mean & 0.966803 \\
SF mean & 0.787312 \\
DF mean & 0.532005 \\
DP--DF Gap & 0.434798 \\
AUC$_{\mathrm{DP-DF}}$ & 1.000000 \\
AUC$_{\mathrm{DP-SF}}$ & 0.896934 \\
AUC$_{\mathrm{DP-All}}$ & 0.983232 \\
Mean parent margin & 0.110509 \\
\bottomrule
\end{tabular*}
\caption{Final point estimates on the released 411-pair benchmark.}
\label{tab:final_bref_point_estimates}
\end{table}

Suspect-clustered and family-clustered uncertainty estimates are summarized in Table~\ref{tab:clustered_ci}. Paired MRR bootstrap and Holm-corrected exact McNemar tests are reported in Table~\ref{tab:stats_main}; all five baseline comparisons remain significant at $p<0.05$ after correction.

\begin{table*}[!t]
\centering
\small
\setlength{\tabcolsep}{5pt}
\renewcommand{\arraystretch}{1.16}
\begin{tabular*}{\textwidth}{@{\extracolsep{\fill}}lrrr@{}}
\toprule
Metric & Point estimate & Suspect-clustered 95\% CI & Family-clustered 95\% CI \\
\midrule
DP mean & 0.966803 & [0.954075, 0.978284] & [0.951799, 0.980552] \\
DF mean & 0.532005 & [0.523167, 0.540695] & [0.519287, 0.547560] \\
DP--DF Gap & 0.434798 & [0.422463, 0.447537] & [0.418080, 0.448046] \\
AUC$_{\mathrm{DP-DF}}$ & 1.000000 & [1.000000, 1.000000] & [1.000000, 1.000000] \\
AUC$_{\mathrm{DP-All}}$ & 0.983232 & [0.973601, 0.993309] & [0.971070, 0.995775] \\
Mean parent margin & 0.110509 & [0.052629, 0.177008] & [0.053891, 0.215700] \\
\bottomrule
\end{tabular*}
\caption{Clustered uncertainty for the main BReF statistics.}
\label{tab:clustered_ci}
\end{table*}

\begin{table*}[h]
\centering
\small
\setlength{\tabcolsep}{6pt}
\renewcommand{\arraystretch}{1.16}
\begin{tabular*}{\textwidth}{@{\extracolsep{\fill}}lrrr@{}}
\toprule
Comparison & Full $\Delta$MRR & Family-bootstrap 95\% CI & LOFO $\Delta$MRR range \\
\midrule
BReF $-$ REEF & 0.1829 & [0.0625, 0.2647] & [0.1190, 0.2118] \\
BReF $-$ MET & 0.3038 & [0.1153, 0.4782] & [0.2372, 0.3518] \\
BReF $-$ ZeroPrint & 0.5766 & [0.4530, 0.7123] & [0.5452, 0.6225] \\
BReF $-$ LLMmap & 0.5896 & [0.5003, 0.7312] & [0.5595, 0.6403] \\
BReF $-$ QA-Agreement & 0.2798 & [0.0952, 0.4799] & [0.1907, 0.3239] \\
\bottomrule
\end{tabular*}
\caption{Family-level sensitivity of BReF's paired MRR advantage over all five baselines. Family-bootstrap intervals use 10,000 family-clustered replicates; LOFO reports the range across six leave-one-family-out evaluations.}
\label{tab:family_mrr_sensitivity}
\end{table*}

\subsection{Family-level dependence sensitivity}
The primary uncertainty analysis clusters by suspect, which accounts for the multiple candidate comparisons attached to each query model. Residual dependence can still remain among suspects from the same model family. The 22 suspects form six family clusters: Qwen (8), Llama-lineage (4), DeepSeek (3), Mistral (3), Baichuan (2), and Falcon (2). We therefore repeat the bootstrap by resampling model families as blocks. This analysis relaxes the independence assumption; it does not claim that family-level resampling establishes independence.

The paired MRR advantage of BReF remains positive against all five baselines under family-level resampling. Table~\ref{tab:family_mrr_sensitivity} reports the full-sample $\Delta$MRR, the family-clustered 95\% confidence interval, and the range obtained from six leave-one-family-out (LOFO) evaluations. Every family-bootstrap interval lies strictly above zero, and removing any one family preserves a positive MRR advantage. For QA-Agreement, the family-clustered interval is $[0.0952,0.4799]$ and the LOFO range is $[0.1907,0.3239]$; excluding Qwen alone still leaves $\Delta$MRR $=0.2968$. These analyses are treated as conservative sensitivity checks because only six family clusters are available.

\section{Reproducibility Checklist}
The experiment archive contains the benchmark definitions, model and
provenance mappings, frozen probe pool and clean13 transformations, cached
pair scores and ranks, reproduced baselines, statistical and robustness
analyses, scripts, logs, cache mappings, and a SHA256 manifest. Package
validation confirms 411 main pairs, 22 suspects, 22 provenance edges, and
34 public repositories.

\section{Ethical Considerations}
Model fingerprinting can support provenance auditing and intellectual-property investigation, but a behavioral similarity score can also be misused for vendor de-anonymization or unsupported accusations. BReF should be treated as technical evidence rather than standalone proof of ownership or misconduct. Attribution conclusions should be combined with public lineage documentation, licenses, release chronology, metadata, and human review.

\end{document}